\UseRawInputEncoding
\documentclass[superscriptaddress,12pt]{revtex4}
\usepackage{amsfonts}
\usepackage{graphicx, amsmath}
\usepackage{subfigure}
\begin{document}
\title[Short Title]{Modified quantum delayed-choice experiment without quantum control}
\author{Qi Guo\footnote{E-mail: qguo@sxu.edu.cn}}
\affiliation{College of Physics and Electronic Engineering, Shanxi University, Taiyuan, Shanxi 030006, People's Republic of China}
\affiliation{State Key Laboratory of Quantum Optics and Quantum Optics Devices, and Institute of Opto-Electronics, Shanxi University, Taiyuan 030006, China}
\affiliation{Collaborative Innovation Center of Extreme Optics, Shanxi University, Taiyuan 030006, China}
\author{Wen-Jie Zhang}
\affiliation{College of Physics and Electronic Engineering, Shanxi University, Taiyuan, Shanxi 030006, People's Republic of China}
\affiliation{State Key Laboratory of Quantum Optics and Quantum Optics Devices, and Institute of Opto-Electronics, Shanxi University, Taiyuan 030006, China}
\affiliation{Collaborative Innovation Center of Extreme Optics, Shanxi University, Taiyuan 030006, China}
\author{Gang Li}
\affiliation{State Key Laboratory of Quantum Optics and Quantum Optics Devices, and Institute of Opto-Electronics, Shanxi University, Taiyuan 030006, China}
\affiliation{Collaborative Innovation Center of Extreme Optics, Shanxi University, Taiyuan 030006, China}
\author{Tiancai Zhang\footnote{E-mail: tczhang@sxu.edu.cn}}
\affiliation{State Key Laboratory of Quantum Optics and Quantum Optics Devices, and Institute of Opto-Electronics, Shanxi University, Taiyuan 030006, China}
\affiliation{Collaborative Innovation Center of Extreme Optics, Shanxi University, Taiyuan 030006, China}
\author{Hong-Fu Wang}
\affiliation{Department of Physics, College of Science, Yanbian University, Yanji, Jilin 133002, People's Republic of China}
\author{Shou Zhang}
\affiliation{Department of Physics, College of Science, Yanbian University, Yanji, Jilin 133002, People's Republic of China}
\begin{abstract}

Wheeler's delayed-choice experiment delays the decision to observe either the wave or particle behavior of a photon until after it has entered the interferometer, and the quantum delayed-choice experiment provides the possibility of observing the wave and particle behavior simultaneously by introducing quantum control device. We here propose a modified quantum delayed-choice experiment without quantum control or entanglement assistance, in which a photon can be prepared in a wave-particle superposition state and the morphing behavior of wave-to-particle transition can be observed easily. It is demonstrated that the presented scheme can allow us to rule out classical hidden variable models in a device-independent manner via violating dimension witness. We also extend the scheme to the situation of two degrees of freedom, first constructing a hybrid quantum delayed-choice experiment which enables simultaneous observation of a photon's wave and particle behaviors in different degrees of freedom, and then proposing a scheme to prepare the single-photon wave-particle entanglement. This study is not only meaningful to explore the wave and particle properties of photons, but also provides potential for the research of the single-particle nonlocality from the perspective of the wave-particle degree of freedom.

\end{abstract}
\maketitle

\section{Introduction}

Wave-particle duality, one of the most fascinating characters of quantum physics, means a quantum object has both wave-like and particle-like properties that are two distinct and mutually exclusive natures from the perspective of classical physics. Bohr's complementarity principle shows that the wave behavior and particle behavior cannot be observed simultaneously, and which behavior a quantum object will exhibit depends on the measurement arrangement \cite{1}. Mach-Zehnder interferometer (MZI) provides an effective platform for testing the wave-particle duality of a single photon. The fist beam splitter (BS) of a MZI splits the input photon into two paths, and the second BS of the MZI recombine the two paths. Therefore, if the second BS is inserted into the MZI, the interference between the two paths can be observed at the two output ports, and the input photon shows wave behavior; if the second BS is removed, the which-path information will be revealed at the output ports, and the photon shows particle behavior.

However, one objection is that maybe a hidden variable in the initial state can tell the input photon in advance the second BS is inserted or not, so the photon can adjust its behavior to the corresponding measurement apparatus. To rule out the hidden variable theory, Wheeler proposed the famous delay-choice experiment \cite{2,3}, in which the second BS is decided to be inserted or removed after the photon has entered the interferometer, so the photon cannot know which measurement apparatus lie ahead in advance. With the development of experiment technology, this Gedanken experiment has been realized in actual laboratory by using different system, such as photons \cite{4,5,6}, and atoms \cite{7,8,9}, even been implemented between satellite and ground stations \cite{10}. Moreover, the delay-choice experiment was also extended to other domains of quantum physics \cite{9,11,12,13,14,15,16,17}, such as delay-choice quantum erase \cite{12,13}, delayed-choice entanglement swapping \cite{14,15,16}, delayed-choice decoherence suppression \cite{17}, entanglement-separation duality \cite{16}, and so on. These experiments exhibit profound and amazing quantum effects. On the other hand, the wave-particle duality has also been studied quantitatively by the complementarity inequality \cite{18,19,20}.

The reason that the wave behavior and particle behavior of a photon cannot be observed simultaneously is the two measurement apparatuses (removing the second BS or not) are mutually exclusive. However, a quantum version delayed-choice experiment was proposed by replacing the second BS in MZI with a quantum-controlled BS \cite{21}, in which the second BS can be prepared in a quantum superposition state of presence and absence by using an ancilla qubit. Thus, the quantum delayed-choice (QDC) experiment enables the simultaneous observation of a photon's wave and particle behaviors. The key module of the QDC experiment is the quantum control device of the second BS, therefore, many researchers have designed the quantum BS for either massless photons or massive particles to implement the QDC experiment in different systems \cite{22,23,24,25,26}, and observed the wave-particle superposition behavior. The QDC experiment enriches the basic contents of wave-particle duality and Bohr's complementarity principle, so people's interest in the foundations of quantum mechanics has been further stimulated, and a variety of works exploring quantum phenomena under the frame of the QDC experiment has been proposed in theory \cite{27,28,29,30} and experiment \cite{31,32,33,34,35,36,37,38}.

The original intention of Wheeler's delay-choice (WDC) experiment is to exclude classical hidden variable models. However, very recently, Chaves \emph{et al.} considered the WDC experiment and its quantum version from the perspective of device-independent causal models \cite{39}, and proposed that a causal two-dimensional hidden-variable theory can reproduce the quantum mechanical predictions of the WDC experiment and the QDC experiment, which means the original WDC experiment cannot rule out the classical hidden variable model. In that work, the authors treated the WDC experiment as a device-independent prepare-and-measure (PAM) scenario \cite{40}, and suggested a slight modification for the WDC experiment can exclude any two-dimensional nonretrocausal hidden classical model in a device-independent manner by violating the dimension witness inequality \cite{40,41,42}. Subsequently, this causal-modeled WDC experiment was carried out independently in a series of experiments \cite{43,44,45}. The causal modeling approach \cite{46} provides an effective way to analyze the nonclassical nature of an experiment by classical causal assumptions \cite{47,48,49}.  These theoretical and experimental works gave further evidences for the nonclassicality of wave-particle duality.

The essential advance of the QDC experiment over the WDC experiment is that the wave behavior and particle behavior of a quantum object can be observed simultaneously due to the quantum-controlled BS, i.e., the wave-particle superposition state can be prepared. In the existing schemes for the QDC experiment, the quantum control device were achieved with the help of ancilla qubits  \cite{21,22,23,24,25,26} or entanglement \cite{27,35}. Here, we first propose an alternative proposal for QDC experiment with classical strategies, that is, the photon's wave-particle superposition state can be generated without any quantum control or entanglement assistance, and the same statistical result as the existing schemes can be obtained. We demonstrate the theoretical results of the presented scheme can violate the dimension witness, which means it can rule out classical hidden variable models in a device-independent manner. Moreover, we generalize this proposal to the case of two degrees of freedom, which shows that a single photon can behave as wave and particle simultaneously in the different degrees of freedom, and the single-photon wave-particle entanglement can also be prepared.

\section{The quantum delayed-choice experiment with tunable beam splitter}

We now introduce how to realize the QDC experiment with classical strategies. The schematic depiction of the presented proposal is shown in Fig.~1(a). Note that, actually, the complete setup diagram of the presented proposal is Fig.~1(c). Fig.~1(a) can only generate the same statistics results as Fig.~1(c), but cannot rule out the two-dimensional hidden-variable model in Ref.~\cite{39}, which can be ruled out in Fig.~1(c) by inserting an additional phase shifter as shown in the following section. Here, to exhibit the photon's statistics clearly, we explain the procedure using Fig.~1(a). The second BS in the MZI is replaced by a tunable beam splitter (TBS) with reflectivity $\cos^{2}\theta$ and transmissivity $\sin^{2}\theta$, where $\theta$ is continuously tunable between 0 and $\frac{\pi}{4}$. The TBS is always placed in the interferometer in the proposed scheme (the idea of delayed choice will reflect in the choice of $\phi$ in Fig.~1(c)). We denote the two paths of the MZI with quantum states $|0\rangle$ and $|1\rangle$. Let a photon enter the MZI initially from the path 0, i.e., the initial state is $|0\rangle$. The BS transforms the state as $\{|0\rangle\rightarrow\frac{1}{\sqrt{2}}(|0\rangle+|1\rangle),~|1\rangle\rightarrow\frac{1}{\sqrt{2}}(|0\rangle-|1\rangle)\}$, and the phase shifter $\varphi$ induces a phase shift $\varphi$ for the photon in the path 1. Thus, after the photon passing through BS and $\varphi$, the state evolves to
\begin{eqnarray}\label{e1}
|\psi\rangle\rightarrow\frac{1}{\sqrt{2}}(|0\rangle+e^{i\varphi}|1\rangle).
\end{eqnarray}
Then the photon is reflected by mirrors (MR) and reaches the TBS, whose action is equivalent to the rotation $\{|0\rangle\rightarrow\cos\theta|0\rangle+\sin\theta|1\rangle,~|1\rangle\rightarrow\sin\theta|0\rangle-\cos\theta|1\rangle\}$. When the photon leaves the MZI, the state becomes
\begin{eqnarray}\label{e2}
|\psi\rangle_{f}=\frac{1}{\sqrt{2}}[(\cos\theta+e^{i\varphi}\sin\theta)|0\rangle+(\sin\theta-e^{i\varphi}\cos\theta)|1\rangle].
\end{eqnarray}

\begin{figure}
\renewcommand\figurename{\small Fig.}
 \centering \vspace*{8pt} \setlength{\baselineskip}{10pt}
 \subfigure[]{
 \includegraphics[scale = 1.0]{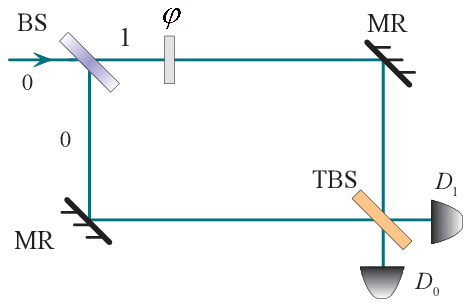}}
 \subfigure[]{
 \includegraphics[scale = 0.8]{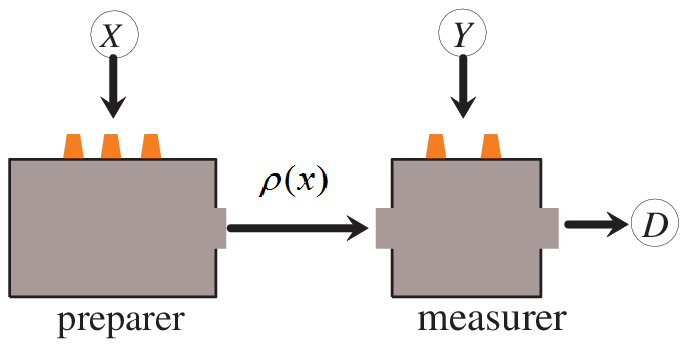}}
 \subfigure[]{
 \includegraphics[scale = 1.0]{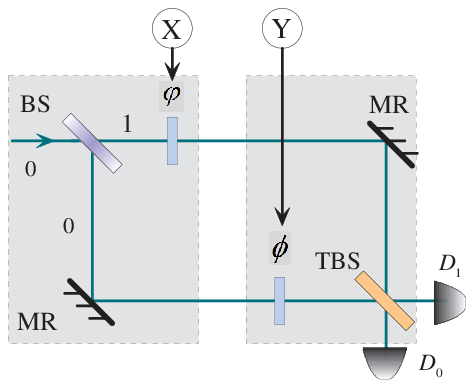}}
 \caption{\label{f1} (a) The QDC experiment without quantum control device. MR: normal mirror. BS: 50:50 beam splitter. $\varphi$: phase shifter. TBS: tunable beam splitter. The two paths of the MZI are labeled by 0 and 1. $D_{0(1)}$: conventional photon detector. (b) The device-independent PAM scenario consists of a preparer (the first black box) with buttons $X$, a measurer (the second black box) with buttons $Y$, and a detection $D$. (c) The complete schematic of the proposed QDC experiment with an additional phase shifter $\phi$ that can be used to exclude the two-dimensional hidden-variable model. The gray dashed boxes correspond to the preparer and the measurer in (b).}
\end{figure}

Obviously, if $\theta=0$, $|\psi\rangle_{f}=\frac{1}{\sqrt{2}}(|0\rangle-e^{i\varphi}|1\rangle)$, detectors $D_{0(1)}$ can reveal the which-path information of the photon in the MZI, and the photon behaves as a particle; if $\theta=\frac{\pi}{4}$, $|\psi\rangle_{f}=e^{i\varphi/2}(\cos\frac{\varphi}{2}|0\rangle-i\sin\frac{\varphi}{2}|1\rangle)$ the photon behaves as a wave. Therefore, following the operational description of the wave and particle behavior of a photon in Ref.~\cite{21}, we can introduce the definition of the particle state $|\mathrm{particle}\rangle=\frac{1}{\sqrt{2}}(|0\rangle-e^{i\varphi}|1\rangle)$ and the wave state $|\mathrm{wave}\rangle=e^{i\varphi/2}(\cos\frac{\varphi}{2}|0\rangle-i\sin\frac{\varphi}{2}|1\rangle)$. This operational definition not only provides suitable expressions for the capacity and incapacity of the photon to produce interference in the context of quantum mechanics, but also can be conveniently used to study the intermediate behavior and the transition behavior between wave and particle nature \cite{21,22,23,24,25,26,27,35}. In the wave-particle representation, the final state in Eq.~(2) can be rewritten as
\begin{eqnarray}\label{e3}
|\psi\rangle_{f}=\alpha|\mathrm{particle}\rangle+\beta|\mathrm{wave}\rangle,
\end{eqnarray}
where the coefficients $\alpha=\cos\theta-\sin\theta$ and $\beta=\sqrt{2}\sin\theta$. It can be seen from Eq.~(3), for $\theta=0$ and $\frac{\pi}{4}$, the photon is in the particle and wave state respectively. While $\theta$ is an arbitrary value between 0 and $\frac{\pi}{4}$, Eq.~(3) is a wave-particle superposition state, which means the photon will exhibit wave property and particle property simultaneously. In order to show the intermediate morphing behavior between wave and particle nature visually, we should explore the interference pattern at the output ports of the MZI, which, for a single photon, can be reflected by the probabilities that the detector $D_{0(1)}$ clicks. Take the output port 0 for example, the probability of $D_{0}$ clicking is
\begin{eqnarray}\label{e4}
I(\varphi,\theta)=\mathrm{Tr}(\rho_{f}|0\rangle\langle0|)=\frac{1}{2}(1+\sin2\theta\cos\varphi),
\end{eqnarray}
where $\rho_{f}=|\psi\rangle_{ff}\langle\psi|$ is the density matrix of the photon's final state. For an arbitrary $\theta$, the visibility of the interference pattern at the output port can be obtained $V=(I_{max}-I_{min})/(I_{max}+I_{min})=\sin2\theta$. We plot the probability distribution $I(\varphi,\theta)$ versus $\theta$ and $\varphi$ in Fig.~2, from which one can see the continuously morphing behavior between wave ($\theta=\pi/4$) and particle ($\theta=0$). Thus, by varying $\theta$ we can observe the photon's behavior of wave-to-particle transition, that is to say, both wave and particle properties can be measured in a single experiment by  classical strategies and without quantum control.

\begin{figure}
\scalebox{0.6}{\includegraphics{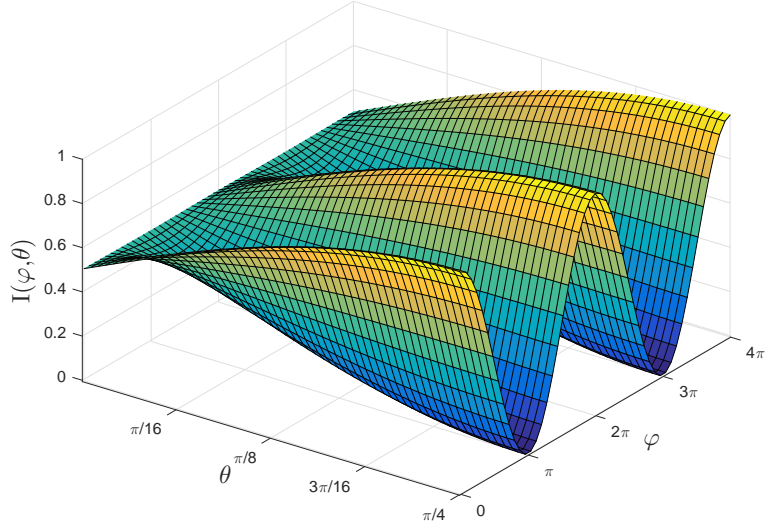}}\caption{\label{f2}
The morphing behavior between wave ($\theta=\pi/4$) and particle ($\theta=0$) by continuously tuning $\theta$.}
\end{figure}

\section{Ruling out the classical hidden variable models via dimension witness method}

The complementary properties can be observed in a single experimental setup with the presented scheme, but the setup above cannot rule out the classical hidden variable that maybe tell the input photon in advance about the value of $\theta$. One possible way is to tune the parameter $\theta$ of the TBS when the photon has entered the MZI and before it reaches the TBS, so that the photon cannot adjust itself beforehand to the specific superposition state corresponding to $\theta$. However, the same as the WDC experiment, this method can be explained by the classical two-dimensional hidden variable causal model proposed in Ref.~\cite{39}. To exclude the the causal model, we here adopt the method similar to Ref.~\cite{39}, which can be demonstrated with the device-independent PAM scenario by the violation of dimension witness. The PAM scenario as shown in Fig.~1(b) consists of a preparer (the first black box), a measurer (the second black box), a detector $D$ \cite{40,41}. The preparer can prepare a physical system in the state $\rho(x)$ by pressing one of the buttons $X$, and then the system is sent to the measurer. By choosing one of the buttons $Y$, the system is measured and an outcome $D$ will be produced. For the PAM scenario, to produce the same statistical distribution, a classical system is required higher dimensionality than a quantum system, which is the theoretical basis for testing classical and quantum systems by using the dimension witness. For example, consider a scenario with $2k$ preparations and $k$ binary measurements, the dimension witness can be achieved with the help of the $k\times k$ matrix \cite{40}
\begin{eqnarray}\label{e5}
\mathrm{W}_{k}(i,j)=p(2j,i)-p(2j+1,i),
\end{eqnarray}
with $0\leq i,j\leq k-1$, and $p(x,y)=p(D=0|x,y)$. The dimension witness $|\mathrm{Det}(\mathrm{W}_{k})|$ equals 0 for any classical system of dimension $d\leq k$, but for any quantum system of dimension $d\leq \sqrt{k}$. Hence we can test classical and quantum systems by using the dimension witness $|\mathrm{Det}(\mathrm{W}_{k})|$.

In order to rule out the classical causal model, we should insert an extra phase shifter $\phi$ in the path 0 as shown in Fig.~1(c), which has the same experimental results and photon statistical behaviors as Fig.~1(a) by absorbing $\phi$ into $\varphi$, so it can also measure the photon's wave property and particle property at the same time. As demonstrated by Chaves \emph{et al.} \cite{39}, the delayed-choice experiment is equivalent to the PAM scenario. The first dotted rectangular box and the second one in Fig.~1(c) correspond to the preparer and the measurer in the PAM scenario. The value of $\phi$ should be chosen after the preparation process to ensure there is no correlation between the preparer and the measurer. We also sent the photon from path 0 initially, and after passing through the BS and $\varphi$, the quantum state corresponding to $\rho(x)$ is $\frac{1}{\sqrt{2}}(|0\rangle+e^{i\varphi}|1\rangle)$. Then the photon enters the measurer, and passes through $\phi$ and TBS. The state evolves to
\begin{eqnarray}\label{e6}
|\psi'\rangle_{f}=\frac{1}{\sqrt{2}}[(e^{i\phi}\cos\theta+e^{i\varphi}\sin\theta)|0\rangle+(e^{i\phi}\sin\theta-e^{i\varphi}\cos\theta)|1\rangle].
\end{eqnarray}

Because there are two detection results or the photon has two spatial modes in the experiment, the dimension of system to be test is 2. Therefore, we should set four preparation choices $X(\mathrm{i.e.} \varphi)\in\{\varphi_{0},\varphi_{1},\varphi_{2},\varphi_{3}\}$, and two measurement choices $Y(\mathrm{i.e.} \phi)\in\{\phi_{0},\phi_{1}\}$. The matrix used for dimension witness is given by
\begin{eqnarray}\label{e7}
\mathrm{W}_{2}=\left(\begin{array}{cc}
p(0,0)-p(1,0)&p(2,0)-p(3,0)\\
p(0,1)-p(1,1)&p(2,1)-p(3,1)\\
\end{array}
\right),
\end{eqnarray}
where
$p(x,y)=p(D=0|\varphi_{x},\phi_{y})$ is the probability the photon is detected by $D_{0}$ for the choice $\varphi_{x}$ and $\phi_{y}$. For the state in Eq.~(6), it can be obtained that $p(x,y)=\frac{1}{2}[1+\sin2\theta\cos(\varphi_{x}-\phi_{y})]$. We can get
\begin{eqnarray}\label{e8}
\mathrm{W}_{2}=\frac{1}{4}\sin^{2}2\theta\{&[\cos(\varphi_{0}-\phi_{0})-\cos(\varphi_{1}-\phi_{0})][\cos(\varphi_{2}-\phi_{1})-\cos(\varphi_{3}-\phi_{1})]
\cr\cr&-[\cos(\varphi_{2}-\phi_{0})-\cos(\varphi_{3}-\phi_{0})][\cos(\varphi_{0}-\phi_{1})-\cos(\varphi_{1}-\phi_{1})\},
\end{eqnarray}
Without loss of generality, we choose $\varphi_{0}=\frac{1}{2}\varphi_{1}=\frac{1}{3}\varphi_{2}=\frac{1}{4}\varphi_{3}=\varphi$, $\phi_{0}=0$, and $\phi_{1}=\frac{\pi}{2}$ for evaluating $|\mathrm{Det}(\mathrm{W}_{k})|$, then,
\begin{eqnarray}\label{e9}
\mathrm{W}_{2}=\frac{1}{4}\sin^{2}2\theta(2\sin2\varphi-\sin\varphi-\sin3\varphi).
\end{eqnarray}
Now we plot the change of $|\mathrm{Det}(\mathrm{W}_{2})|$ versus $\varphi$ and $\theta$ in Fig.~3(a). We can see $|\mathrm{Det}(\mathrm{W}_{2})|>0$ in several ares, i.e. the dimension witness is violated. For example, $|\mathrm{Det}(\mathrm{W}_{2})|=0.29$ for $\varphi=3\pi/4$ and $\theta=\pi/5$.

\begin{figure}
\renewcommand\figurename{\small Fig.}
 \centering \vspace*{8pt} \setlength{\baselineskip}{10pt}
 \subfigure[]{
 \includegraphics[scale = 0.39]{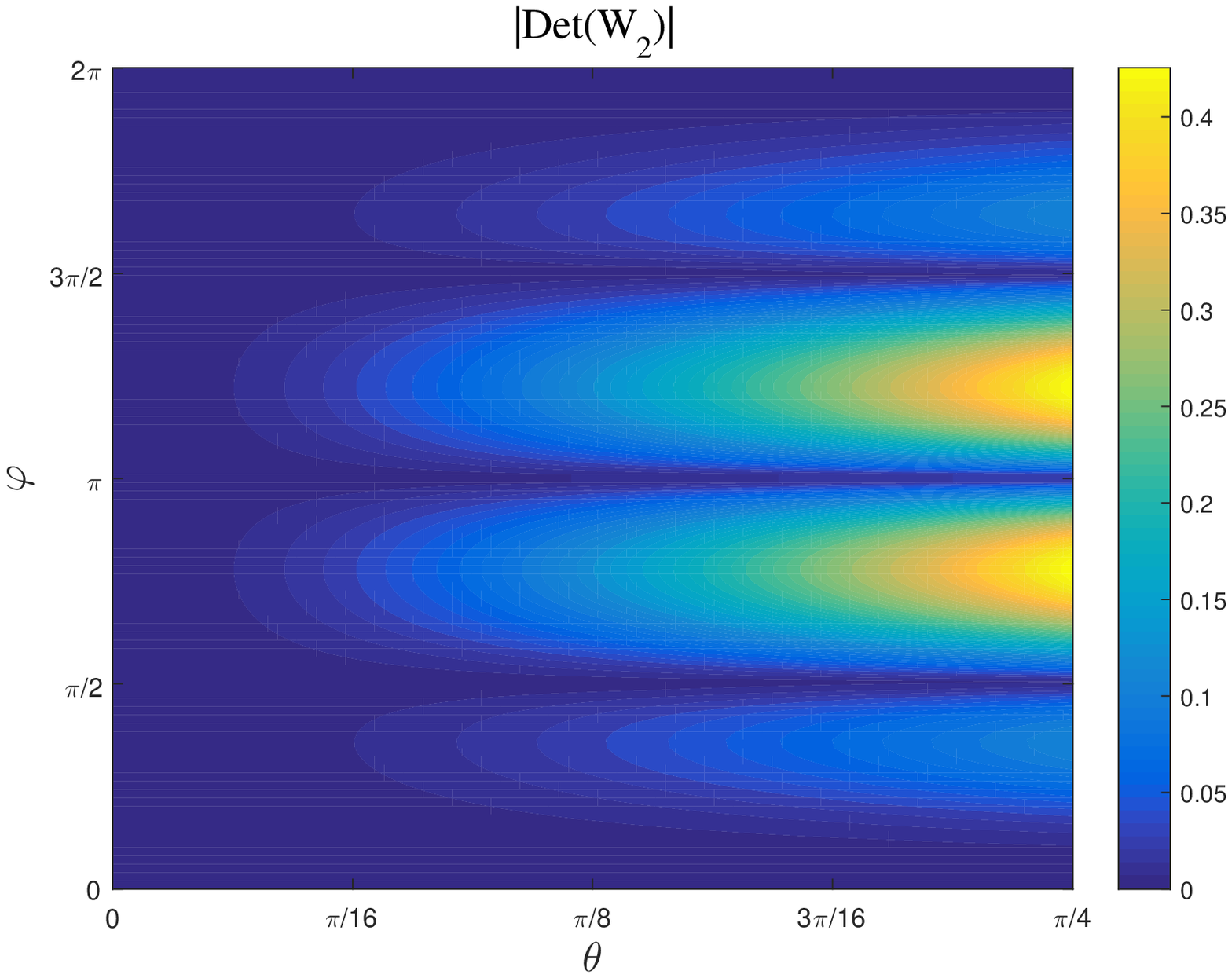}}
 \subfigure[]{
 \includegraphics[scale = 0.39]{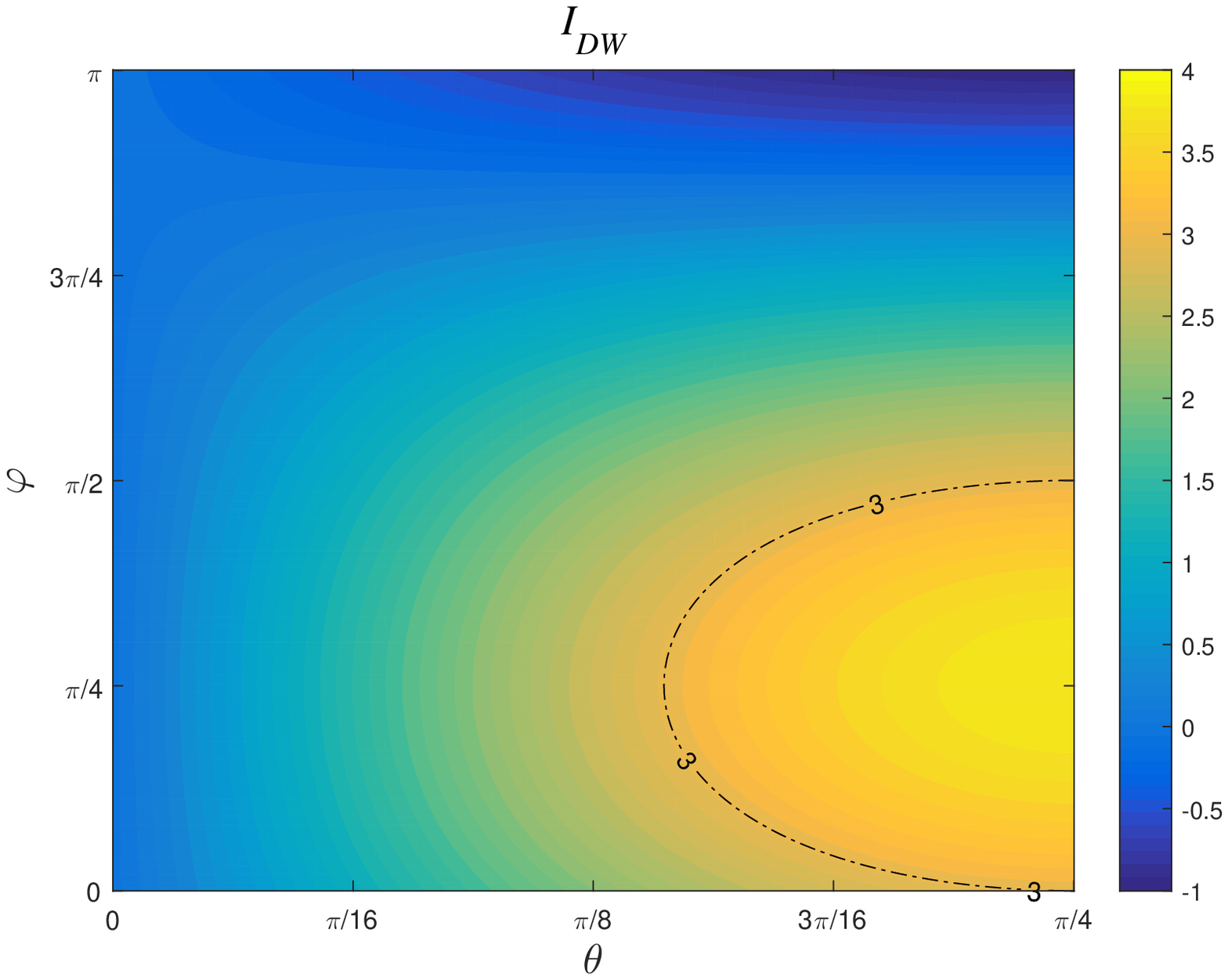}}
  \caption{\label{f3} (a) The dimension witness $|\mathrm{Det}(\mathrm{W}_{2})|$ versus $\varphi$ and $\theta$. (b) The linear dimension witness versus $\varphi$ and $\theta$, and the dimension witness inequality $I_{DW}\leq3$ is violated in the area marked by the dotted line}
\end{figure}

The dimension witness above is called nonlinear dimension witness \cite{40}, that's because it requires the preparer and the measurer are not correlated. Therefore, as pointed out in Ref.~\cite{39}, the analysis above has implicitly assumed that all noise terms are independent and the hidden variable is also independent of any noise term. However, in the causal model proposed in Ref.~\cite{39}, noise terms might affect the hidden variable, and thus influence the statistical behaviors of the photon at the output ports. In order to rule out the hidden variable model with such correlation, we should test the presented scheme with the linear dimension witness, which has been given by an inequality in Ref.~\cite{41}
\begin{eqnarray}\label{e10}
I_{DW}=\langle D_{00}\rangle+\langle D_{01}\rangle+\langle D_{10}\rangle-\langle D_{11}\rangle-\langle D_{20}\rangle\leq3,
\end{eqnarray}
where $\langle D_{xy}\rangle=p(D=0|\varphi_{x},\phi_{y})-p(D=1|\varphi_{x},\phi_{y})$, i.e. the probability difference between $D_{0}$ and $D_{1}$ to detect the photon. To employ the dimension witness inequality, we should set three preparation choices $\varphi\in\{\varphi_{0},\varphi_{1},\varphi_{2}\}$, and two measurement choices $\phi\in\{\phi_{0},\phi_{1}\}$. By using Eq.~(6), we can obtain
\begin{eqnarray}\label{e11}
I_{DW}=\sin2\theta[\cos(\varphi_{0}-\phi_{0})+\cos(\varphi_{0}-\phi_{1})+\cos(\varphi_{1}-\phi_{0})-\cos(\varphi_{1}-\phi_{1})-\cos(\varphi_{2}-\phi_{0})].
\end{eqnarray}
To evaluate $I_{DW}$, we here also choose $\varphi_{0}=-\varphi_{1}=\varphi$, $\phi_{0}=0$, $\phi_{1}=\frac{\pi}{2}$, and $\varphi_{2}=\pi$ for maximizing $I_{DW}$, then $I_{DW}=\sin2\theta[2(\cos\varphi+\sin\varphi)+1]$. From Fig.~3(b), it can be seen that the dimension witness inequality is violated in the area marked by the dotted line, i.e., $I_{DW}>3$ in that region of $(\theta, \varphi)$. Specially, the maximum violation \cite{39,41} in quantum systems can be obtained $I_{DW}=1+2\sqrt{2}$ for $\varphi=\pi/4$ and $\theta=\pi/4$. Thus, the hidden variable correlated with noise terms can also be ruled out in the presented scheme.

\section{Hybrid quantum delayed-choice experiment in different degrees of freedom}

\begin{figure}
\scalebox{1.5}{\includegraphics{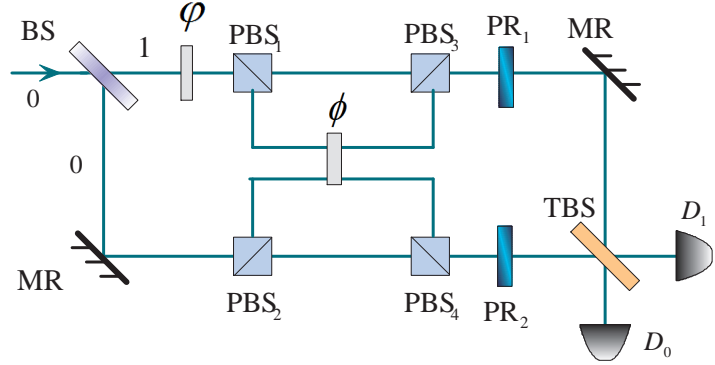}}\caption{\label{f4}
The schematic of the hybrid QDC experiment in path degree of freedom and polarization degree of freedom. PBS: polarizing beam splitter. PR: polarization rotator. Other optical elements are the same as Fig.~1.}
\end{figure}

Now we extend the scheme to two degrees of freedom, i.e., implement the QDC experiment simultaneously in path degree of freedom and polarization degree of freedom. The basic setup diagram of the scheme is shown in Fig.~4. The photon is initially prepared in the superposition state of horizontal polarization $|H\rangle$ and vertical polarization $|V\rangle$, and enters the setup from the path 0, that is, the initial state of the photon can be given by $|\Phi_{0}\rangle=\frac{1}{\sqrt{2}}(|H\rangle+|V\rangle)|0\rangle$. The polarization beam splitter (PBS) transmits $|H\rangle$ component and reflects $|V\rangle$ component. The phase shifters $\varphi$ and $\phi$ induces phase shifts $\varphi$ and $\phi$ for $|V\rangle$ component and $|1\rangle$ component, respectively. So after passing through BS, PBSs, and phase shifters, the photon is in the state
\begin{eqnarray}\label{e12}
|\Phi\rangle=\frac{1}{2}(|H\rangle+e^{i\phi}|V\rangle)(|0\rangle+e^{i\varphi}|1\rangle).
\end{eqnarray}
The polarization rotator (PR) is used to rotate the photon by an angle $\vartheta$, i.e., $|H\rangle\rightarrow\cos\vartheta|H\rangle+\sin\vartheta|V\rangle$ and $|V\rangle\rightarrow\sin\vartheta|H\rangle-\cos\vartheta|V\rangle$. The action of TBS with parameter $\theta$ is the same as that in Fig.~1. Therefore, the state after TBS evolves to
\begin{eqnarray}\label{e13}
|\Phi\rangle_{f}=\frac{1}{2}&[(\cos\vartheta+e^{i\phi}\sin\vartheta)|H\rangle+(\sin\vartheta-e^{i\phi}\cos\vartheta)|V\rangle]
\cr&\otimes[(\cos\theta+e^{i\varphi}\sin\theta)|0\rangle+(\sin\theta-e^{i\varphi}\cos\theta)|1\rangle].
\end{eqnarray}
The particle state and the wave state in the polarization degree of freedom can be defined as $|\mathrm{particle}\rangle=\frac{1}{\sqrt{2}}(|H\rangle-e^{i\phi}|V\rangle)$ and $|\mathrm{wave}\rangle=e^{i\phi/2}(\cos\frac{\phi}{2}|H\rangle-i\sin\frac{\phi}{2}|V\rangle)$. Thus, the state in Eq.~(13) is a superposition state of wave and particle in two degrees of freedom. Especially, $|\Phi\rangle_{f}=|\mathrm{particle}\rangle_{pol}|\mathrm{wave}\rangle_{path}$ for $\vartheta=0$ and $\theta=\pi/4$, and $|\Phi\rangle_{f}=|\mathrm{wave}\rangle_{pol}|\mathrm{particle}\rangle_{path}$ for $\vartheta=\pi/4$ and $\theta=0$, where the subscripts $pol$ and $path$ respectively indicate polarization and path degree of freedom. That is, the hybrid QDC experiment allows a single photon to be in particle state in a degree of freedom but in wave state in the other degree of freedom at the same time. Note that we here have used only one phase shifter in each degree of freedom for simplicity, which cannot rule out the causal model. To achieve this, every phase shifter should be divided into two parts and placed in the preparer and the measurer similar to Fig.~1(c), then the causal model can be excluded by using the same way as above section.

\begin{figure}
\scalebox{1.3}{\includegraphics{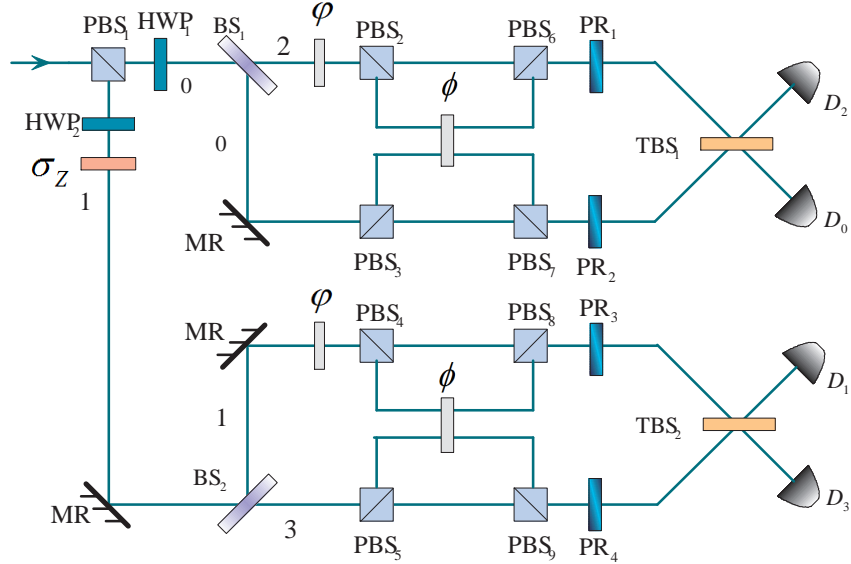}}\caption{\label{f5}
The schematic for generating single photon wave-particle entangled state. HWP: half-wave plate oriented at $22.5^{\circ}$. $\sigma_{z}$: $\pi$-phase shifter Other. optical elements are the same as Fig.~4.}
\end{figure}

The hybrid QDC experiment above can be straightforward used to generate single photon wave-particle entangled state, whose schematic diagram is shown in Fig.~5. Two hybrid QDC experiment setups are combined by one PBS, two half-wave plates (HWP), and a $\pi$-phase shifter $\sigma_{z}$. The photon is initially prepared in the state $|\Psi_{0}\rangle=\frac{1}{\sqrt{2}}(|H\rangle+|V\rangle)|1\rangle$. Through direct calculation, the state of the photon passing through the whole setup becomes
\begin{eqnarray}\label{e14}
|\Psi\rangle_{f}=\frac{1}{2\sqrt{2}}&\{[(\cos\vartheta_{1}|H\rangle+\sin\vartheta_{1}|V\rangle)+e^{i\phi}(\sin\vartheta_{1}|H\rangle-\cos\vartheta_{1}|V\rangle)]
\cr&[(\cos\theta_{1}|0\rangle+\sin\theta_{1}|2\rangle)+e^{i\varphi}(\sin\theta_{1}|0\rangle-\cos\theta_{1}|2\rangle)]
\cr&+[(\cos\vartheta_{2}|H\rangle+\sin\vartheta_{2}|V\rangle)+e^{i\phi}(\sin\vartheta_{2}|H\rangle-\cos\vartheta_{2}|V\rangle)]
\cr&[(\cos\theta_{2}|1\rangle+\sin\theta_{2}|3\rangle)+e^{i\varphi}(\sin\theta_{2}|1\rangle-\cos\theta_{2}|3\rangle)]\},
\end{eqnarray}
where $\vartheta_{1}$ and $\vartheta_{2}$ respectively denote the rotated angles by $\mathrm{PR}_{1(2)}$ and $\mathrm{PR}_{3(4)}$, and $\theta_{1(2)}$ is the transmission parameter of $\mathrm{TBS}_{1(2)}$. When we choose $\vartheta_{1}=\theta_{2}=0$ and $\theta_{1}=\vartheta_{2}=\pi/4$, in the wave-particle representation, the state above can be written as
\begin{eqnarray}\label{e15}
|\Psi\rangle_{f}=\frac{1}{\sqrt{2}}(|\mathrm{particle}\rangle_{pol}|\mathrm{wave}\rangle_{path}+|\mathrm{wave}\rangle_{pol}|\mathrm{particle}\rangle_{path}),
\end{eqnarray}
which is a wave-particle entangled state of a single photon. The concurrence of the state above equals 1, i.e., the state is a maximally entangled state. By choosing proper parameters, other Bell-like states can also be obtained. The entanglement generation scheme can be regarded as a simple application of the presented hybrid QDC experiment. Compared with the two-photon wave-particle entanglement in Ref.~\cite{33}, the single-photon wave-particle entanglement proposed here maybe more counterintuitive for exhibiting the photon's dual wave-particle behavior.

\section{Discussion and conclusions}

The wave-particle duality is a fundamental topic of quantum mechanics. The emergence of the QDC experiment has enriched people's understanding of Bohr's complementarity principle. The relevant QDC schemes presented here requires only the most ordinary optical elements in optical laboratory \cite{50}. Compared to existing schemes, a crucial optical element here is TBS that plays key role in observing the behavior of wave-to-particle transition. Fortunately, such TBS has been realized experimentally \cite{51}. Moreover, the device-independent manner used here is robust to arbitrarily losses inside the interferometer and the inefficiency of detectors as pointed out in Ref.~\cite{39}, which has been demonstrated in current experiments \cite{43,44,45}. Therefore, the presented scheme is feasible under the current experimental condition.

In conclusion, we have proposed an alternative scheme for the QDC experiment without quantum control or entanglement assistance, which means the wave-particle superposition state of a photon can be obtained with classical strategies and provides a compact way to observing the morphing behavior of wave-to-particle transition. By violating nonlinear dimension witness and linear dimension witness inequality, it has been demonstrated that the presented scheme can exclude classical two-dimensional hidden variable causal models in a device-independent manner. We have also constructed a hybrid QDC experiment in two degrees of freedom that makes it possible for a photon to exhibit particle behavior in one degree of freedom but exhibit wave behavior in the other one. The single-photon wave-particle entanglement between two degrees of freedom can be prepared by using the hybrid QDC experiment. Therefore, these works may be meaningful for the research of the single-particle nonlocality and quantum information protocols from the perspective of the wave-particle representation.

\begin{center}$\mathbf{Acknowledgments}$\end{center}

This work is supported by the National Natural Science Foundation of China under Grant No. 11604190 and No. 11974223, the Natural Science Foundation of Shanxi Province No. 201901D211167, Scientific and Technological Innovation Programs of Higher Education Institutions in Shanxi No. 2019L0043, and the Fund for Shanxi ``1331 Project" Key Subjects Construction.

\end{document}